\documentstyle[proceedings,numreferences,epsfig,citesort]{crckapb}

\begin{opening}
\title{
SEGREGATION OF POLYDISPERSE GRANULAR MEDIA IN THE PRESENCE
OF A TEMPERATURE GRADIENT }
\author{STEFAN LUDING, OLIVER STRAUSS,}
\institute{Institute for Computer Applications 1\\
           Pfaffenwaldring 27, 70569 Stuttgart, GERMANY\\
           Correspondence to: \verb|lui@ica1.uni-stuttgart.de| }
\author{SEAN McNAMARA}
\institute{Levich Institute, Steinman Hall T-1M, \\
    140th St and Convent Ave, \\
    New York, NY 10031, USA}

\end{opening}

\runningtitle{SEGREGATION ... TEMPERATURE GRADIENT}

\begin{document}

\def\n{{\mathbf \hat n}}
\def\bv{{\mathbf{v}}}

\begin{abstract}
Granular media are examined with the focus on polydisperse
mixtures in the presence of two localized heat-baths. 
If the two driving energies are similar, the large particles
prefer to stay in the `cold' regions of the system -- as far
away from the energy source as possible. If one of the 
temperatures is larger than the other, the cold region is shifted
towards the colder reservoir; if the temperature of one source is much 
higher, a strong, almost
constant temperature gradient builds up between the two reservoirs 
and the large particles are found close to the cold reservoir.
Furthermore, clustering is observed between the heat
reservoirs, if dissipation is strong enough.
\end{abstract}

\vspace{-1.5cm}~\\

\section{Introduction}

The segregation of granular materials is an effect
of eminent importance for industrial operations and
has been a subject of research for decades.
The behavior of powders within the industrial environment, e.g.
silos, hoppers, conveyor belts or chutes, displays interesting 
effects - one of them is size segregation. Segregation or the 
mixing properties of granular media are not yet completely understood 
and thus cannot be controlled under all circumstances. 
For a review of experimental techniques, theoretical approaches, 
and numerical simulations see Refs.\ \cite{herrmann98,chowhan95} and 
the references therein.

A lot of effort has been invested in the understanding of size-segregation 
(see this proceedings for a recent overview of the state of the art).
It turns out that
segregation can be driven by geometric effects, shear, percolation 
and also by a convective motion of the small particles in the
system \cite{knight93}. Segregation due to convection, in rather dilute,
more dynamic systems appears to be orders of magnitude faster than 
segregation due to purely geometrical effects in dense, quasi-static
situations \cite{duran94,dippel95}. 
However, there are still a lot of open questions which are
subject to current research on model granular media \cite
{rosato87,clement95,cantelaube95,herrmann98,arnarson98}. 

Most of the segregation phenomena are obtained in the presence
of gradients in density or temperature. Here we isolate the latter
case, i.e.~we examine the segregation of two species of grains
in the presence of local heat reservoirs, but in the absence of
external forces like e.g.~gravity.

\section{The inelastic hard sphere model}

In this study, we use the standard interaction model for the
instantaneous collisions of particles with radii $a_k$, and
mass $m_k$, with the subscripts $k=S$ or $L$, for small and 
large particles, respectively.
This model accounts for dissipation, using the restitution coefficient
$r$, and is introduced and discussed in more detail in Refs.\
\cite{jenkins85b,lun91,Goldshtein95,luding98d,arnarson98}.
The post-collisional velocities $\bv'$ are given in terms of 
the pre-collisional velocities $\bv$ by
\begin{eqnarray}
\bv_{1,2}' &=& \bv_{1,2} \mp \frac{(1+r)~m_{\rm red}}{m_{1,2}} \bv_n ~,
\label{eq:collrule}
\end{eqnarray}
with $\bv_n \equiv \left [ (\bv_1 - \bv_2) \cdot \bv_n \right ] \n$, 
the normal component of $\bv_1-\bv_2$, parallel to $\n$, 
the unit vector pointing along the line connecting the 
centers of the colliding particles. The reduced mass is
here $m_{\rm red} = m_1 m_2 / (m_1+m_2)$.
If two particles
collide, their velocities are changed according to Eq.\ (\ref{eq:collrule}).

If a particle $i$ crosses a line of fixed temperature $T_j$, 
its velocity is changed in magnitude, but not in direction,
according to the rule
\begin{equation}
\bv_i' = \pm v^{T_j} \frac{\bv_i}{|\bv_i|} ~,
\label{eq:tj}
\end{equation}
with the random thermal velocity $\pm v^{T_j}$ drawn from a Maxwellian
velocity distribution, and with random sign.  (If the `$+$' sign is
used in Eq.\ (\ref{eq:tj}), a large net mass flux occurs when a cluster of
particles crosses a line of fixed temperature.
If the `$-$' sign is used the two subsystems
have conserved particle numbers, but are still coupled via collisions
across the boundaries.) In 2D one has
$v^{T_j} = \sqrt{ v_x^2 + v_y^2 }~$, where $v_x$ and $v_y$ are the
components of the thermal velocity vector, each distributed according
to a Gaussian distribution. Eq.\ (\ref{eq:tj}) is also applied to
particles after a collision, if the particle's center of mass is
closer than $a_L$ to one of the heat reservoirs $j$.

In order to obtain the two random velocities $v_x$ and
$v_y$, two random numbers $r_1$ and $r_2$, homogeneously distributed 
in the interval $[0:1]$ are used. With the desired typical thermal 
velocity $\bar v = \sqrt{ 2 T_j / m_i }$, one has
\begin{equation}
v_x = \sqrt{- \bar v^2 \ln r_2} \cos( 2 \pi r_1 ) ~, {\rm ~and~ }
\end{equation}
\begin{equation}
v_y = \sqrt{- \bar v^2 \ln r_2} \sin( 2 \pi r_1 ) ~, ~~~~~~~
\end{equation}
using the method by Box and Muller, as described in Ref.\ \cite{press92}.

If the velocity of the particle would be set to ${\bf v}' = (v_x,v_y)$,
artificial peaks in the density at the positions $Z_1$ and $Z_2$ would
be observed. This artefact was the reason to choose the above described
way of thermal coupling. Note, that our coupling to a reservoir does
not guarantee a certain temperature in a small volume around the 
reservoir \cite{cause88,mareschal92}. 
It rather adjusts the velocity of all particles that
came along the reservoir and touched it with their center of mass.
Those particles which approach the reservoir usually have a lower
temperature and thus reduce the mean. As $r$ decreases,
the reduction increases.
Our choice of thermal coupling is arbitrary, however, the discussion 
of thermostating is far from the scope of the present paper -- therefore, 
we restrict ourselves to the method introduced here.

Our method of imposing a temperature gradient does not have any
simple physical analog, but it does allow us to isolate the
effects of the temperature gradient.  More physically feasible
energy sources, such as vibrating walls perturb the motion more than
the method used here.

\section{The event-driven simulation method}

For the simulation of the hard spheres, we use the event-driven 
algorithm, originally introduced by Lubachevsky \cite{lubachevsky91},
and applied to the simulation of granular media e.g.~in Refs.\ 
\cite{luding98d}. 
In these simulations, the particles follow an undisturbed translational
motion until an event occurs. An event is either the collision
of two particles, the crossing of one particle with the boundary
of a cell (in the linked-cell structure which is used for algorithmic 
optimization only), or the crossing of one of
the lines with fixed temperature. A particle-particle collision is
treated as described in the previous section, a cell-boundary 
crossing has no effect on the particle-motion, and the crossing
of a fixed temperature location leads to a change of velocity 
according to Eq.\ (\ref{eq:tj}). Only the particle(s) involved in
the last event is (are) treated and their next event is computed.
In the next step (which does not imply a fixed step in time), the
next event out of all possible events is treated.

\section{Boundary conditions and system parameters}

\begin{figure}[t]
\begin{center}
   $r=1.0$ \hfill $r=0.95$ \\
   \epsfig{file=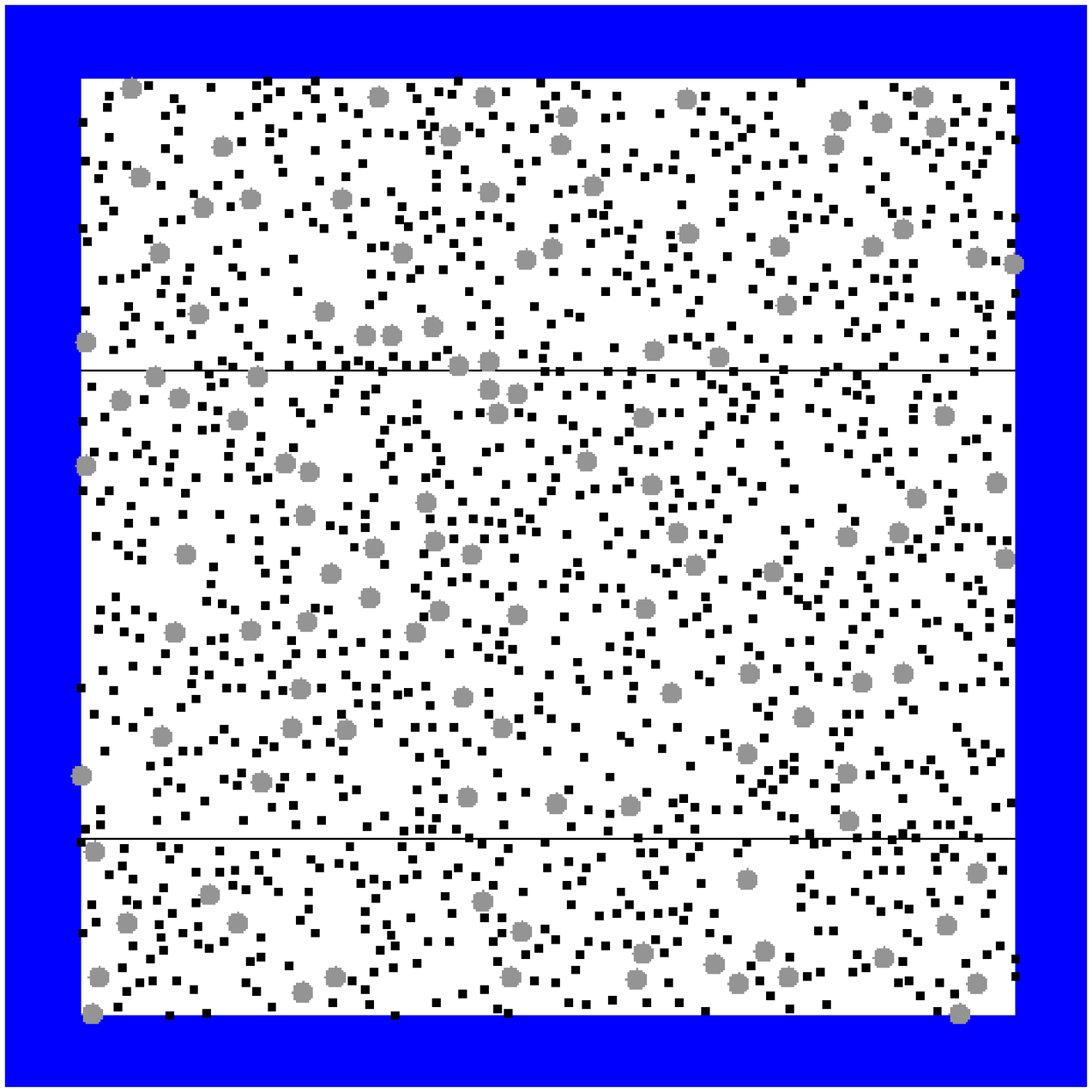,width=6.1cm,clip=,angle=0} \hfill
   \epsfig{file=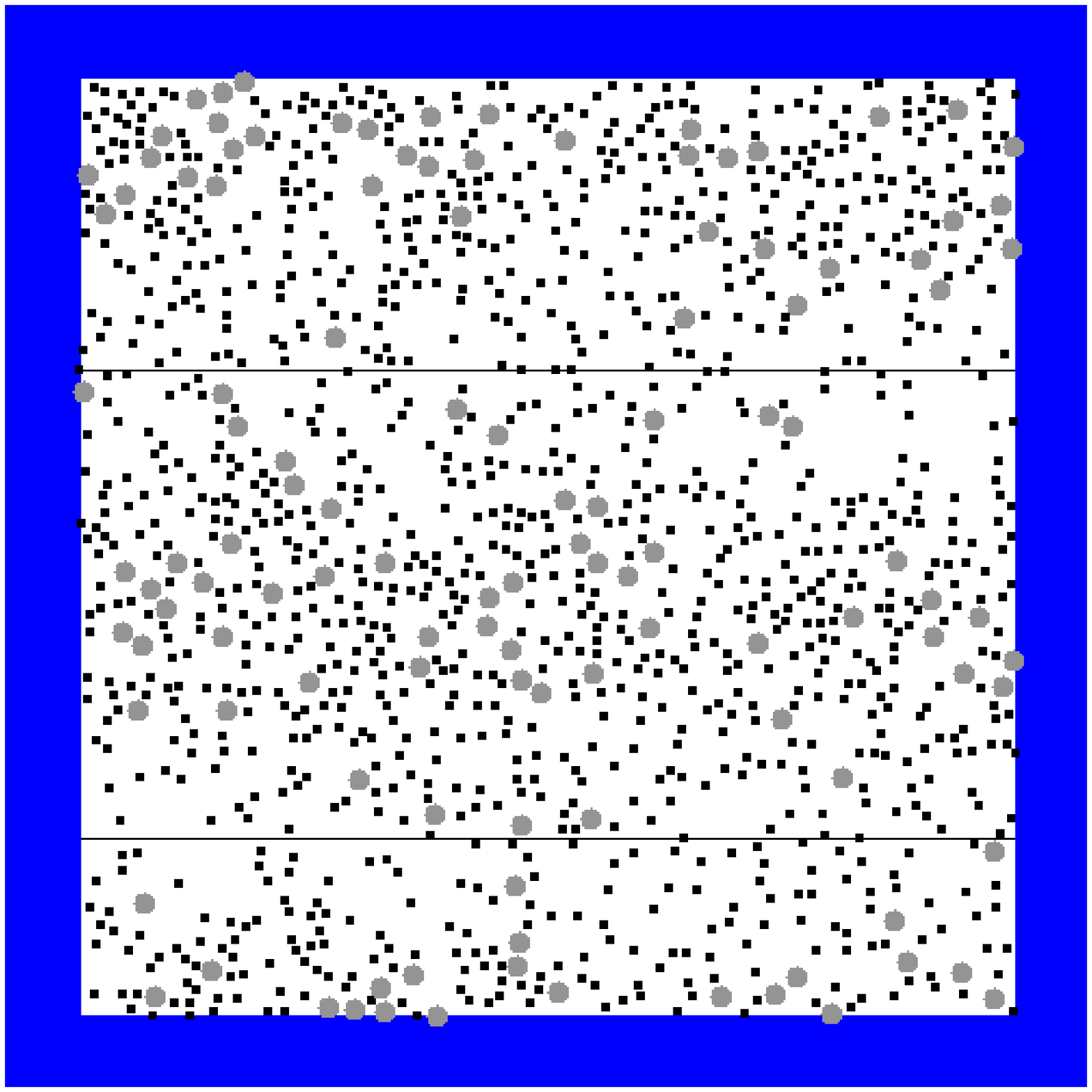,width=6.1cm,clip=,angle=0} \\
 \end{center}
 \begin{center}
   $r=0.90$ \hfill $r=0.60$ \\
   \epsfig{file=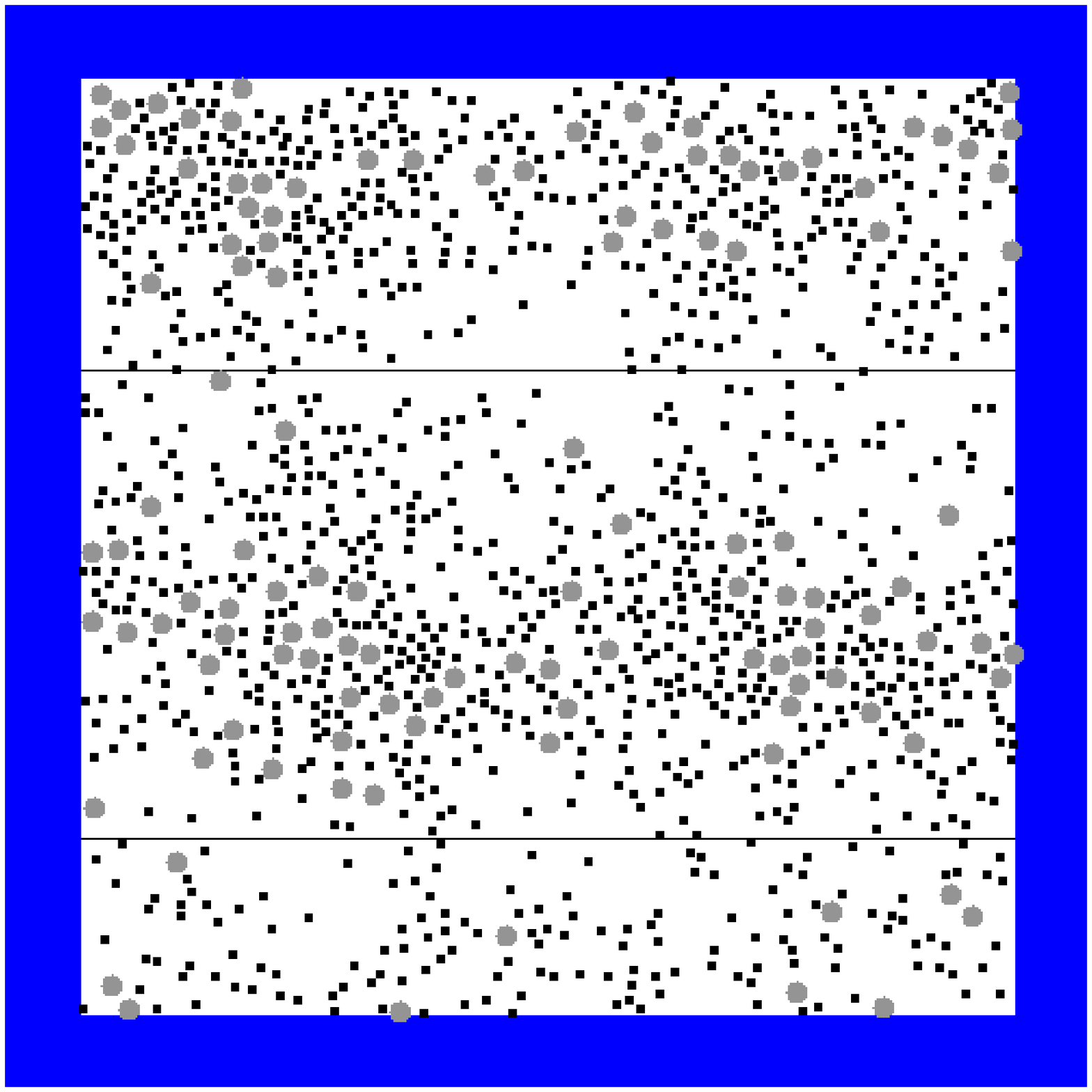,width=6.1cm,clip=,angle=0} \hfill
   \epsfig{file=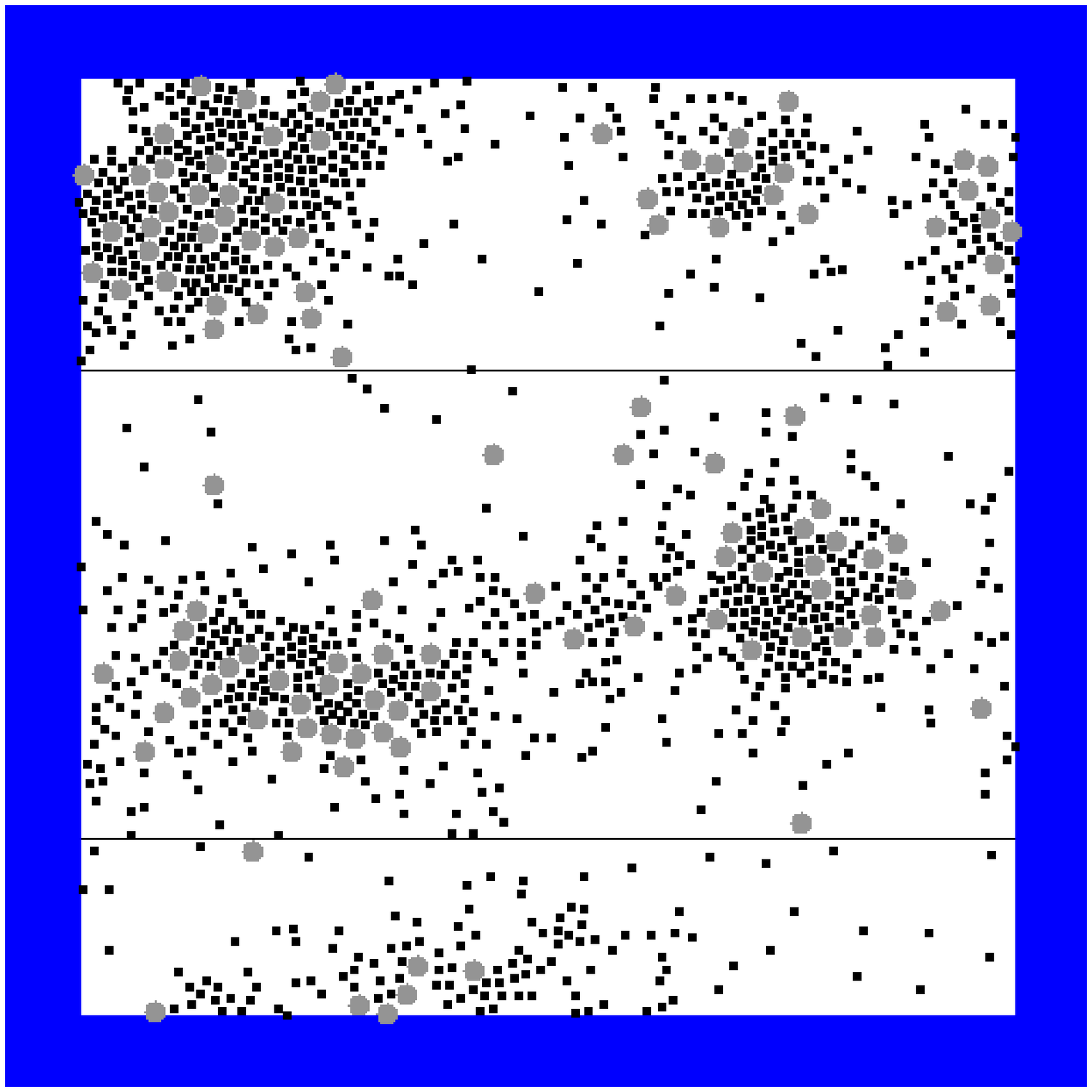,width=6.1cm,clip=,angle=0} \\
\end{center}
\caption{Snapshots from different simulations with $r=1.0$, 0.95, 
0.90, and 0.60. The heat reservoirs are indicated by horizontal lines; 
the large particles are displayed in light grey, the small in black.}
\label{fig:seg0}
\end{figure}

The simulations presented here (typical situations are displayed in 
Fig.\ \ref{fig:seg0}) were performed with $N=1296$ particles
in a two-dimensional (2D) box of size $l=l_x=l_y=0.04$\,m. The box
has periodic boundaries, i.e.~a particle that leaves the volume at the
bottom (left), immediately enters it at the top (right) and vice-versa.
Two particle
types ($S$, $L$) are used with $a_S=a=2.5 \times 10^{-4}$\,m and  $a_L=2a=5
\times 10^{-4}$\,m. In the system with dimensionless size $L=l/a=80$,
$N_S=1176$ particles of the small species and $N_L=120$ particles
of the large species coexist. Note that we used a rather arbitrary
choice for the calculation of the particle mass (see above) so that
$m_L = 8 m_S$ for the size ratio used here. The fraction of the area
which is occupied by the particles, i.e.~the total volume fraction, is
$\nu = \pi (N_S a_S^2 + N_L a_L^2) / l^2 \approx 0.2$.

\begin{figure}[hb]
\begin{center}
 ~\vspace{-0.4cm}\\
 \epsfig{file=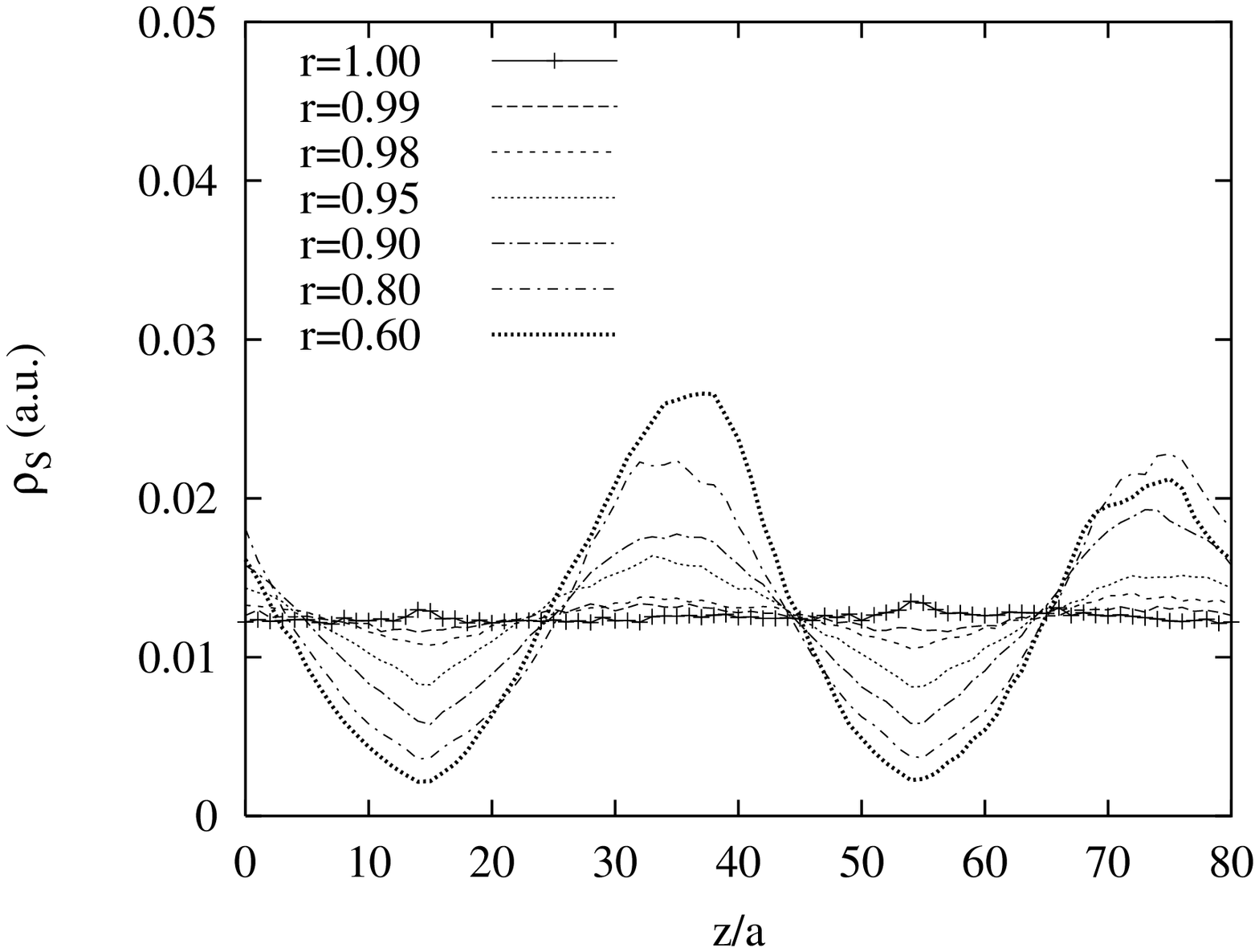,width=4.7cm,clip=,angle=-90}
 \epsfig{file=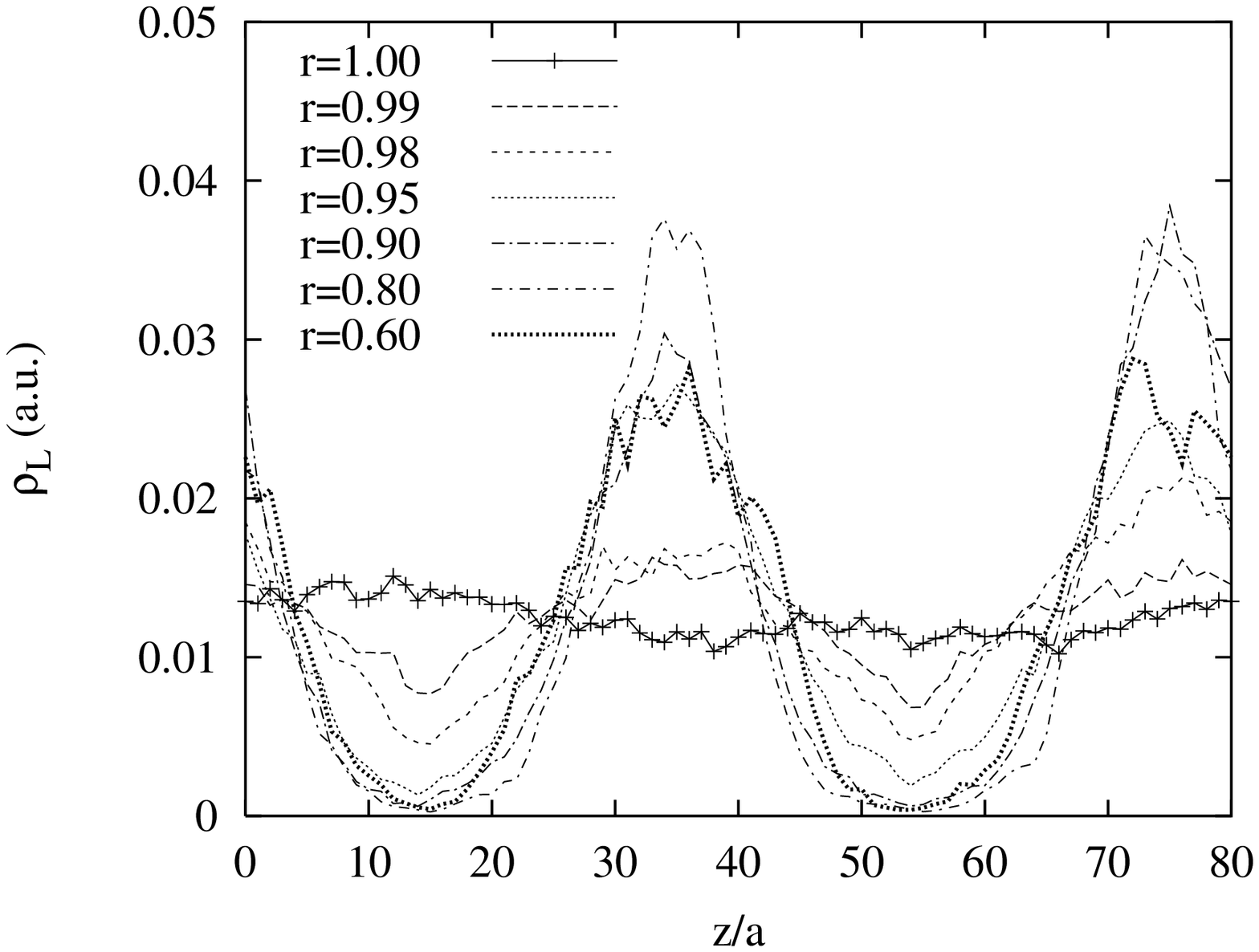,width=4.7cm,clip=,angle=-90}\\
 \epsfig{file=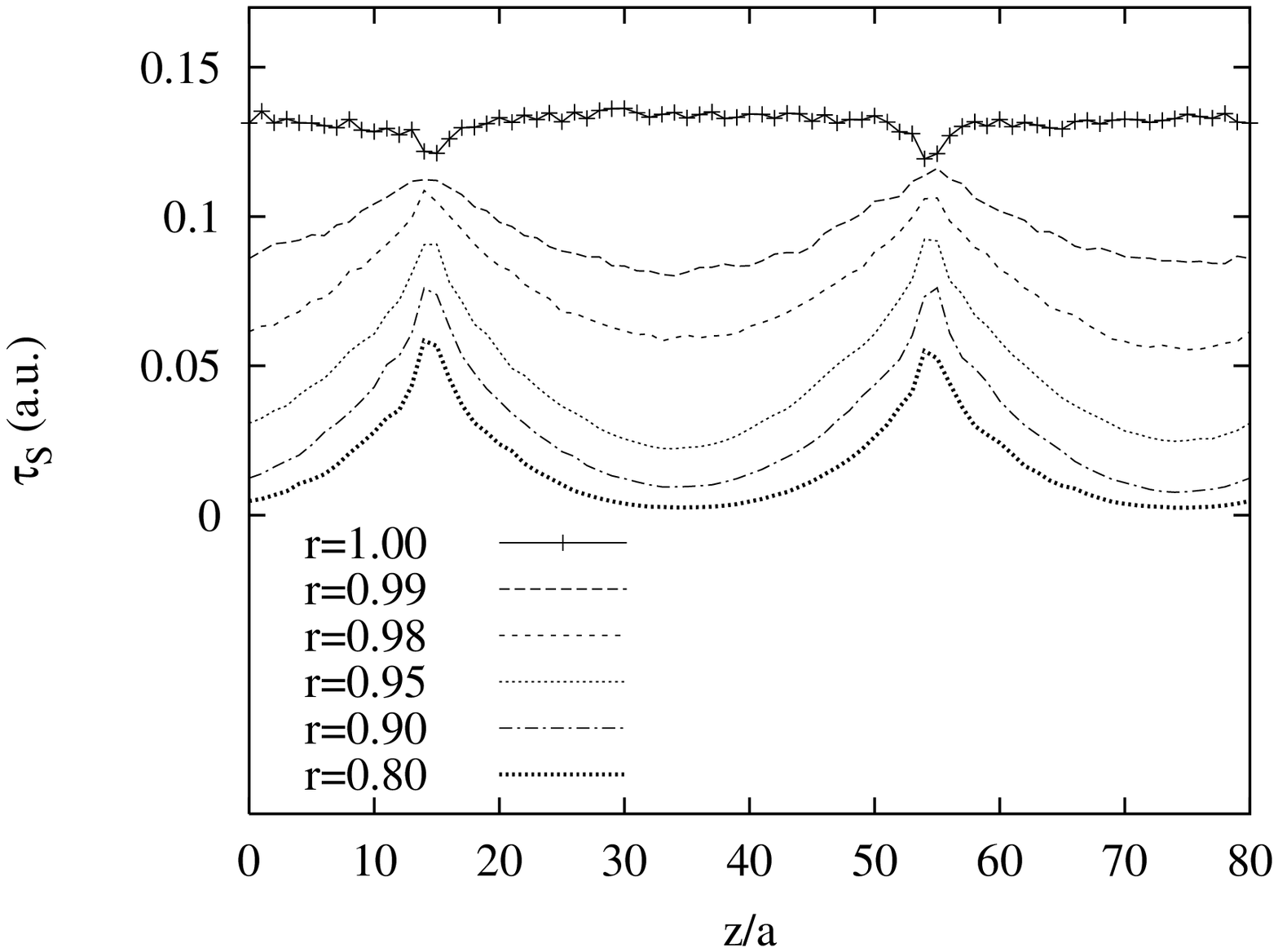,width=4.7cm,clip=,angle=-90}
 \epsfig{file=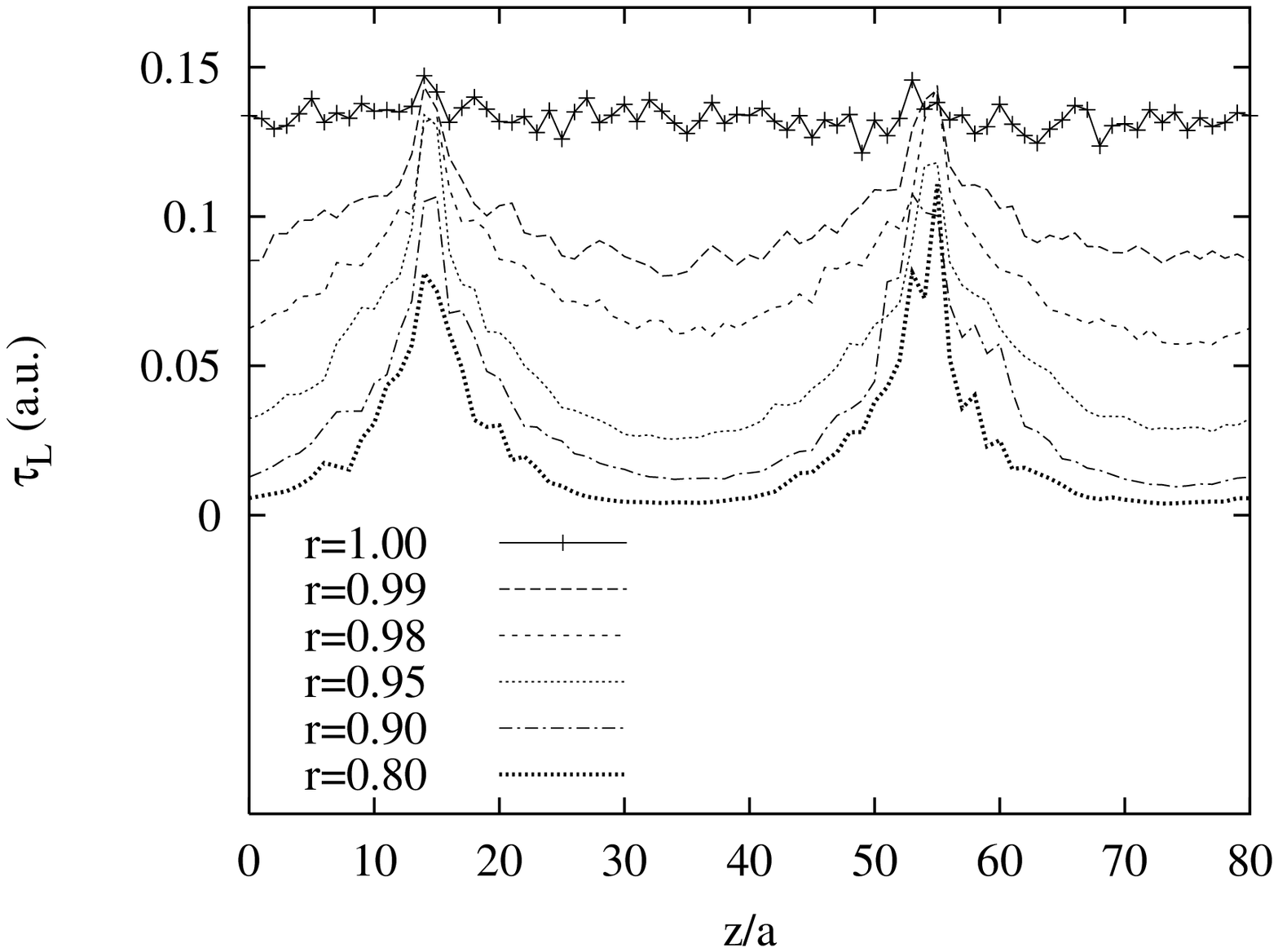,width=4.7cm,clip=,angle=-90} ~\vspace{-0.5cm}\\
\end{center}
\caption{
Density $\rho$ (top) and temperature $\tau$ (bottom),
plotted as function of the vertical position 
in the system $Z=z/a$. Data for small (Left) and large (Right)
particles are presented for $T_2/T_1=1$ and different 
restitution coefficients $r$ as given in the inset. The data at
$z/a=0$ and $z/a=80$ are identical due to periodic boundaries.}
\label{fig:seg1}
\end{figure}

The hot and cold temperature reservoirs are situated at the vertical
positions $Z_2=z_2/a=15$ and $Z_1=z_1/a=55$, respectively, and reach over
the whole width of the system.
The temperatures of the reservoirs are $T_1$ and $T_2$ -- but since
no external body force like gravity is involved the behavior of the
system does not depend on the absolute value of $T_1$ or $T_2$;
only the ratio $T_2/T_1$ is important, and it is varied in the
range from  $T_2/T_1=1$ to 9 in the following. In Fig.\ \ref{fig:seg0} 
snapshots of simulations with $T_2/T_1=1$ and different values of
$r$ are plotted. For small $r$, clustering is observed and the averaging 
over horizontal slices (data are presented in Fig.\ \ref{fig:seg1}) 
becomes questionable.

\section{Results}

A quantitative measure of segregation is the partial
density of the different species. We present $\rho_k = n_k/N_k$,
the particle number density $n_k(z)$ weighed by the number of particles
$N_k$ of species $k$ in Fig.\ \ref{fig:seg1}.
The strength of segregation increases with decreasing $r$, only for 
small $r=0.60$, segregation is rather weak due to clustering.
Fluctuations in density are correlated to the heat reservoirs:
Close to a ``hot'' region in the vicinity of a energy source,
the density is lower than in the ``cold'' regions in between.
Note that the large particles segregate, while
the small particles make up a ``background fluid'' with a
comparatively small density variation, if dissipation is weak.
Thus, all particles prefer regions of low temperature, but the
large ones are attracted by the cold regions more strongly.
\begin{figure}[tb]
\begin{center}
 \epsfig{file=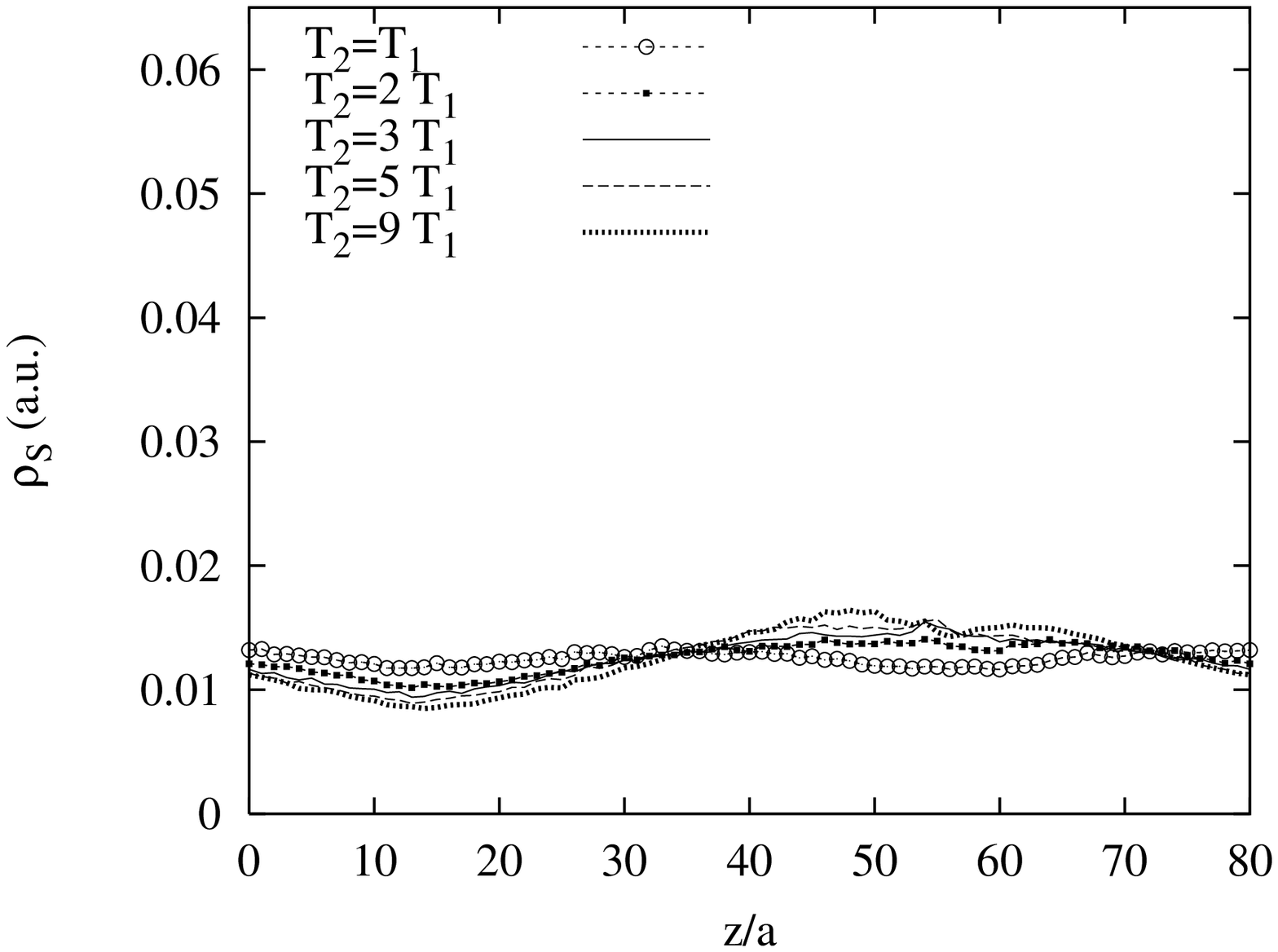,width=4.7cm,clip=,angle=-90}
 \epsfig{file=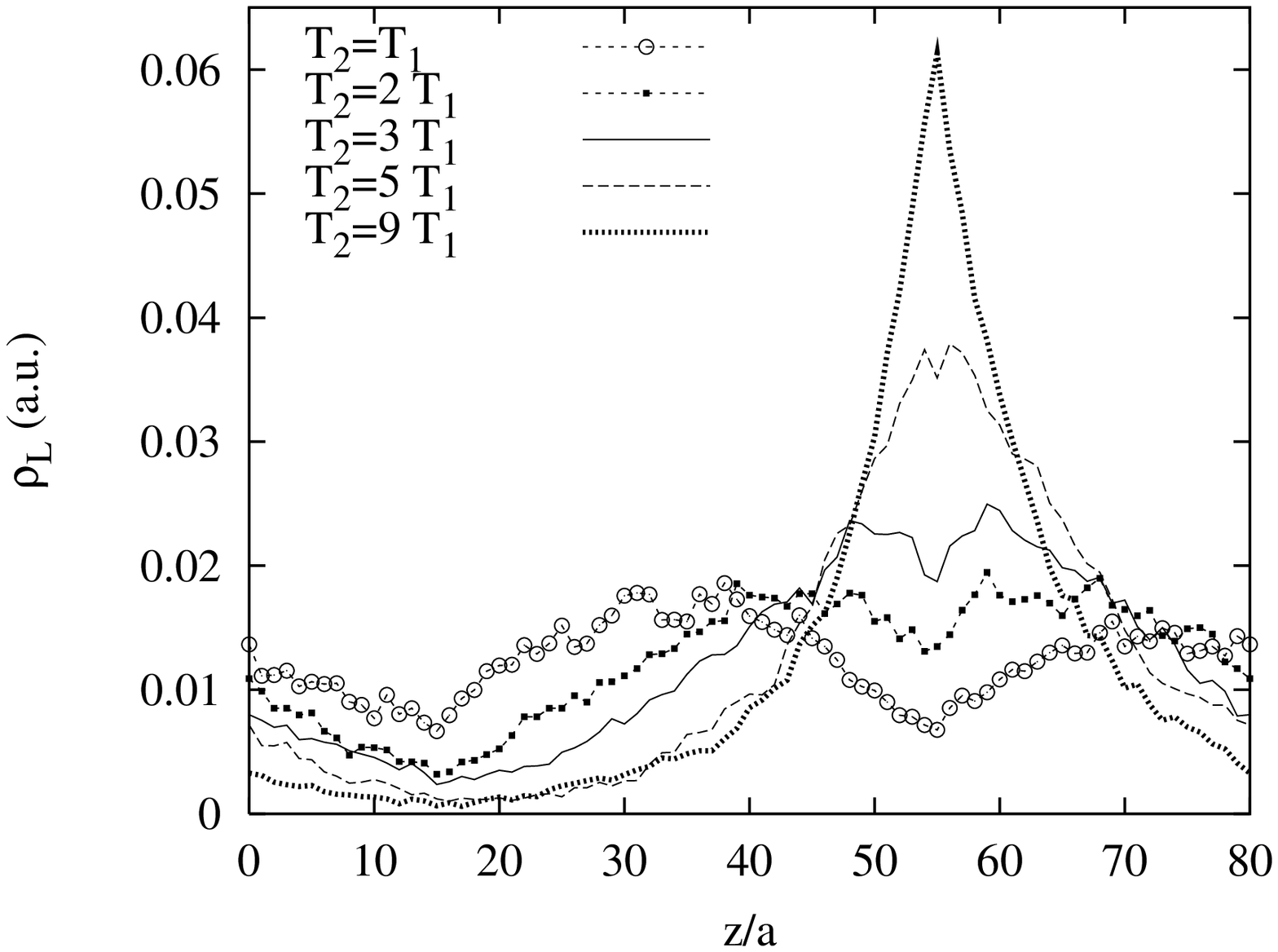,width=4.7cm,clip=,angle=-90}\\
 \epsfig{file=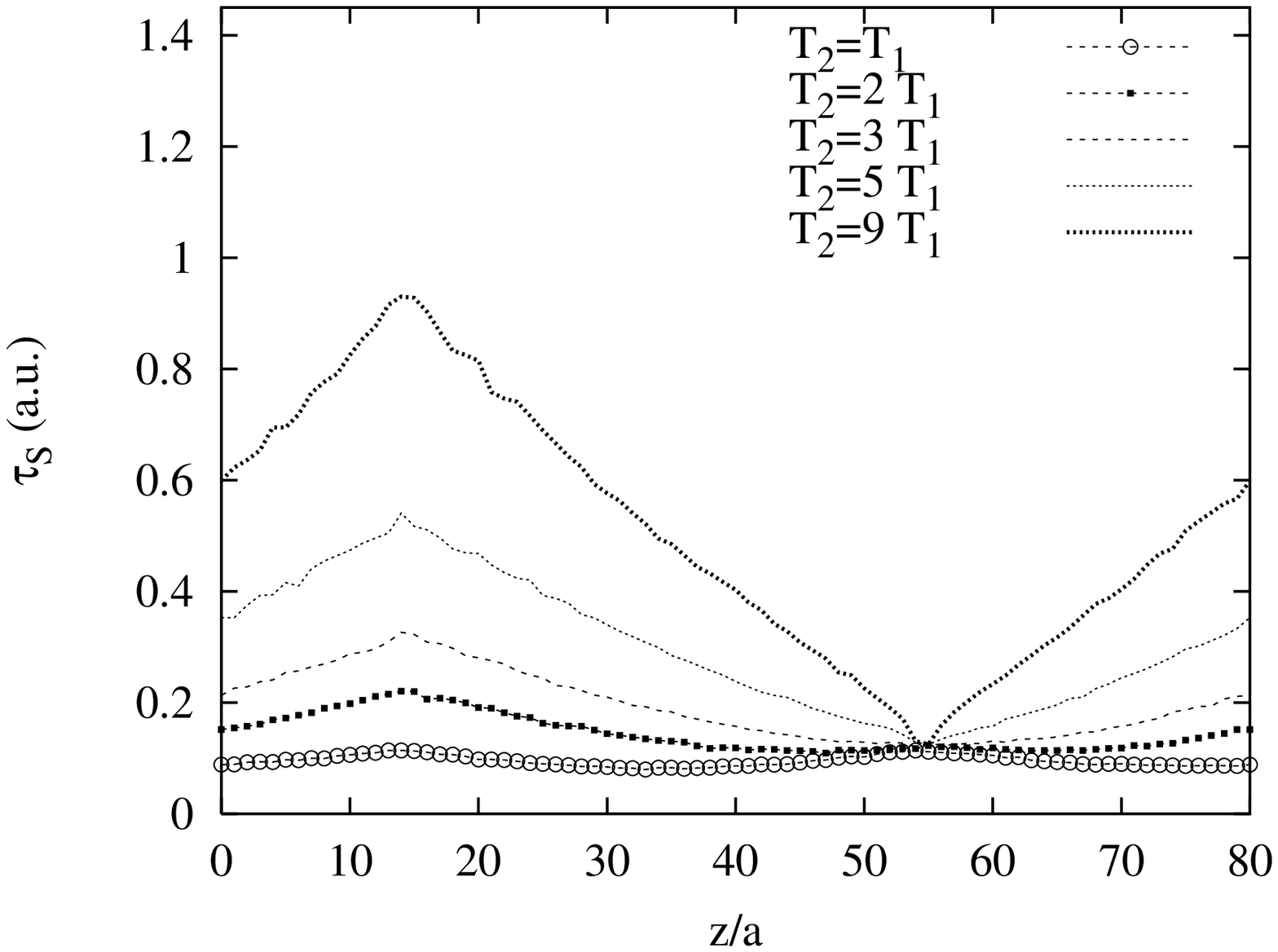,width=4.7cm,clip=,angle=-90}
 \epsfig{file=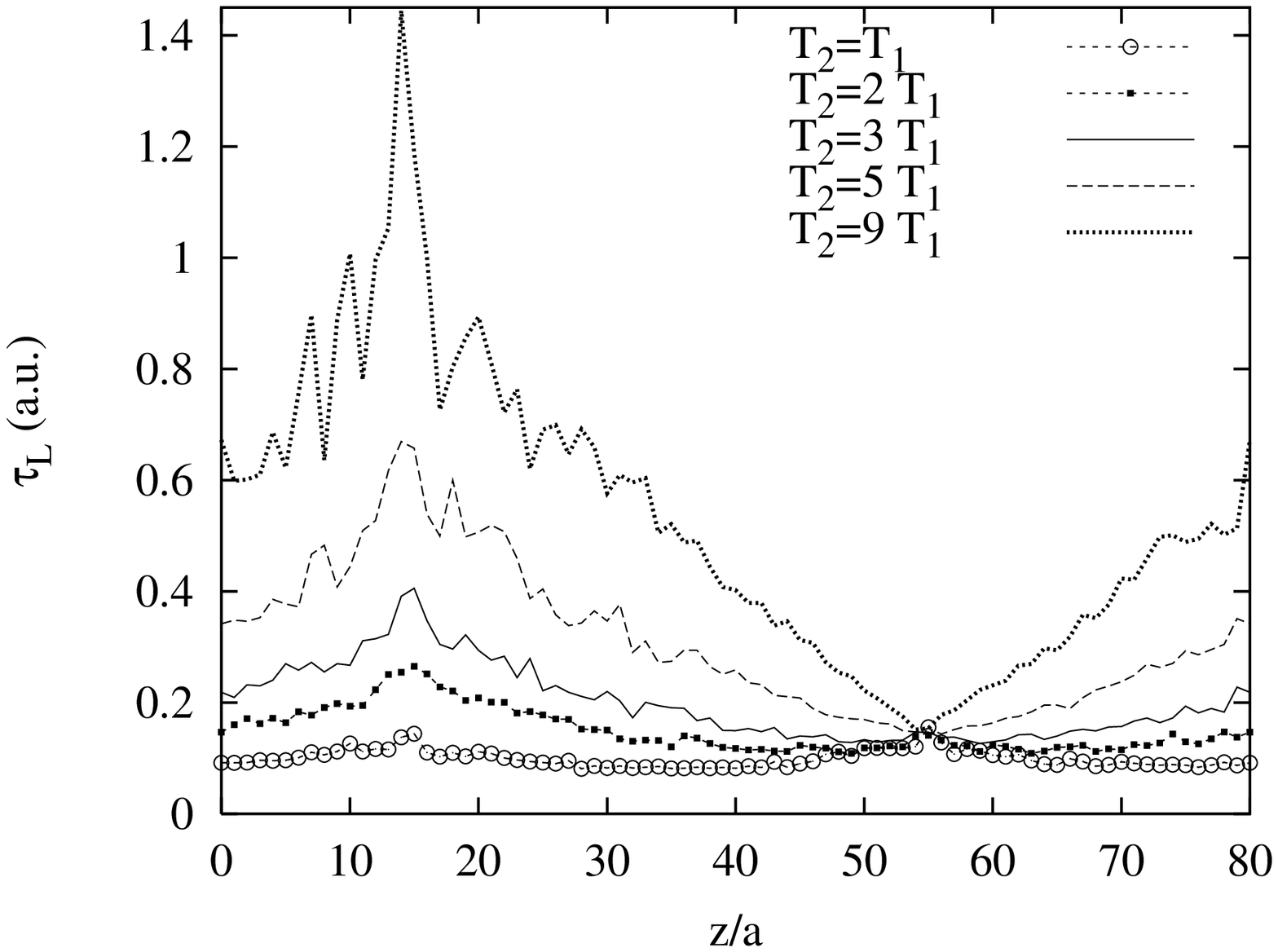,width=4.7cm,clip=,angle=-90} ~\vspace{-0.5cm}\\
\end{center}
\caption{Density $\rho$ (top) and temperature $\tau$ (bottom),
plotted as function of the vertical position
in the system $Z=z/a$. Data for small (Left) and large (Right) particles
are presented for $r=0.99$ and different temperature ratios $T_2/T_1$.}
\label{fig:seg2}
\end{figure}

Now, the restitution coefficient is fixed to $r=0.99$
and the temperature ratio $T_2/T_1$ is modified. 
In Fig.\ \ref{fig:seg2} the situation is
presented for different ratios $T_2/T_1=1$, 2, 3, 5, and 9.
The large particles segregate from the ``background fluid'' made
up by the small particles and the quality of segregation increases
with the magnitude of the temperature gradient.
For $T_2 > T_1$ the large (and heavier) particles can be
found close to the colder heat reservoir, as different to
the situation discussed above. When heavier particles have the
same temperature as light ones, their mean velocity is reduced,
so that they can not diffuse away from the cold heat reservoir.

\section{Summary and Conclusion}

In summary, we presented simulations of different size particles in
the presence of a temperature gradient. If the temperature gradient
is due to the dissipative nature of the material, the large particles
move towards the cold regions, as far away from an energy source as 
possible. If the gradient in temperature is externally imposed, most
of the large particles prefer to move towards the colder heat reservoir.
When dissipation is strong enough, clustering is observed in
the locally driven system.

Further possible studies involve the prediction of the density and
temperature profiles with kinetic theory -- first for the monodisperse
and later also for the polydisperse case.

\section*{Acknowledgements}

We gratefully acknowledge the support of IUTAM, the National Science 
Foundation, the Department of Energy, the Office of Basic Energy Sciences,
the Geosciences Research Program, the Deutsche 
Forschungsgemeinschaft (DFG), and the Alexander-von-Humboldt foundation.


\begin{thebibliography}{10}

\bibitem{herrmann98}
{\em Physics of dry granular media - NATO ASI Series E 350}, edited by H.~J.
  Herrmann, J.-P. Hovi, and S. Luding (Kluwer Academic Publishers, Dordrecht,
  1998).

\bibitem{chowhan95}
Z.~T. Chowhan, Pharm. Technol. {\bf 19},  56  (1995).

\bibitem{knight93}
J.~B. Knight, H.~M. Jaeger, and S.~R. Nagel, Phys. Rev. Lett. {\bf 70},  3728
  (1993).

\bibitem{duran94}
J. Duran, T. Mazozi, E. Cl\'ement, and J. Rajchenbach, Phys. Rev. E {\bf 50},
  5138  (1994).

\bibitem{dippel95}
S. Dippel and S. Luding, J. Phys. I France {\bf 5},  1527  (1995).

\bibitem{rosato87}
A.~D. Rosato, K.~J. Strandburg, F. Prinz, and R.~H. Swendsen, Phys. Rev. Lett.
  {\bf 58},  1038  (1987).

\bibitem{clement95}
E. Cl\'ement, J. Rajchenbach, and J. Duran, Europhys.Lett. {\bf 30},  7
  (1995).

\bibitem{cantelaube95}
F. Cantelaube, Y.~L. Duparcmeur, D. Bideau, and G.~H. Ristow, J. Phys. I France
  {\bf 5},  581  (1995).

\bibitem{arnarson98}
B. Arnarson and J.~T. Willits, Phys. Fluids {\bf 10},  1324  (1998).

\bibitem{jenkins85b}
J.~T. Jenkins and M.~W. Richman, Phys. of Fluids {\bf 28},  3485  (1985).

\bibitem{lun91}
C.~K.~K. Lun, J. Fluid Mech. {\bf 233},  539  (1991).

\bibitem{Goldshtein95}
A. Goldshtein and M. Shapiro, J. Fluid Mech. {\bf 282},  75  (1995).

\bibitem{luding98d}
S. Luding, M. Huthmann, S. McNamara, and A. Zippelius, Phys. Rev. E {\bf 58},
  3416  (1998).

\bibitem{press92}
W.~H. Press, S.~A. Teukolsky, W.~T. Vetterling, and B.~P. Flannery, {\em
  Numerical Recipes} (Cambridge University Press, Cambridge, 1992).

\bibitem{cause88}
P.~J. Cause and M. Mareschal, Phys Rev A {\bf 38}, 4241 (1988).

\bibitem{mareschal92}
M. Mareschal, M. Monsour, G. Sonnino, and E. Kestamont,
 Phys Rev A {\bf 45}, 7180 (1992).

\bibitem{lubachevsky91}
B.~D. Lubachevsky, J. of Comp. Phys. {\bf 94},  255  (1991).

\end{thebibliography}

\end{document}